\documentclass[12pt]{article}
\usepackage{graphicx}
\usepackage{citehack,cite}

\newcommand{\non}{\nonumber \\}
\newcommand{\ve}[1]{{\bf #1}}
\newcommand{\be}{\begin{equation}}
\newcommand{\ee}{\end{equation}}
\newcommand{\bea}{\begin{eqnarray}}
\newcommand{\eea}{\end{eqnarray}}
\newcommand{\sli}{\sum\limits}
\newcommand{\lp}{\left (}
\newcommand{\rp}{\right )}
\newcommand{\lb}{\left \{}

\newcommand{\lbr}{\left [}
\newcommand{\rbr}{\right ]}
\newcommand{\ld}{\left .}
\newcommand{\rd}{\right .}
\newcommand{\rhok}{\rho_{\ve{k}}}
\newcommand{\rhomk}{\rho_{-\ve{k}}}
\newcommand{\cW}{{\cal W}}
\newcommand{\mt}{m_{\tau}}
\newcommand{\np}{n_p}
\newcommand{\Delb}{\Delta_1}
\newcommand{\hhcqp}{\tilde h^2+\tilde h_c^2}
\newcommand{\ccaaHd}{c_{20}^{(0)}H_3}
\newcommand{\btPhia}{\beta\tilde{\Phi}(0)}
\newcommand{\btPhik}{\beta\tilde{\Phi}(k)}
\newcommand{\tPhik}{\tilde{\Phi}(k)}
\newcommand{\tPhia}{\tilde{\Phi}(0)}
\newcommand{\half}{\frac{1}{2}}

\begin{document}

\begin{center}
{\bf THREE-DIMENSIONAL ISING-LIKE SYSTEM IN AN EXTERNAL FIELD:
MICROSCOPIC CALCULATION OF THE FREE ENERGY IN THE HIGHER NON-GAUSSIAN
APPROXIMATION}
\end{center}

\begin{center}
{\sc I.V. Pylyuk}
\end{center}

\begin{center}
{\it Institute for Condensed Matter Physics  \\
of the National Academy of Sciences of Ukraine, \\
1~Svientsitskii Str., UA-79011 Lviv, Ukraine} \\
E-mail: piv@icmp.lviv.ua
\end{center}

\vspace{0.5cm}

{\small
An analytic method for deriving the free energy of a
three-dimensional Ising-like system near the critical point in a
homogeneous external field is developed in the $\rho^6$ model
approximation. The mathematical description proposed for temperatures
$T>T_c$ ($T_c$ is the phase-transition temperature in the absence of
an external field) is valid for fields near $\tilde h_c$, where the
scaling variable is of the order of unity and power series in
this variable are not effective. At the limiting field $\tilde h_c$,
the temperature and field effects on the system in the vicinity of the
critical point are equivalent. The total free energy is obtained as a
function of temperature, field and microscopic parameters of the system
without using series expansions in the scaling variable. In addition to
leading terms, the expression for the free energy includes the terms
determining the temperature and field confluent corrections.
}

\vspace{0.5cm}


\section{Introduction}

In spite of great successes in the investigation of three-dimensional
($3D$) Ising-like systems made by means of various methods (see, for
example, \cite{pv102}), the study of the effect of an external magnetic
field on the critical behaviour of the mentioned systems and the
calculation of scaling functions are still of interest \cite{efs103}.
The Ising model is widely used in the theory of phase transitions for the
study of the properties of various magnetic and non-magnetic systems
(ferroelectrics, ferromagnets, binary mixtures, etc.).

In this article, the behaviour of a $3D$ Ising-like system near the
critical point in a homogeneous external field is studied using the
collective variables (CV) method \cite{ymo287,rev9789,ykpmo101}. The
main peculiarity of this method is the integration of short-wave
spin-density oscillation modes, which is generally done without using
perturbation theory. The CV method is similar to the Wilson
non-perturbative renormalization-group (RG) approach (integration on fast
modes and construction of an effective theory for slow modes)
\cite{btw102,tw194,bb101}. The term collective variables is a common name
for a special class of variables that are specific for each individual
physical system \cite{ymo287,rev9789}. The CV set contains variables
associated with order parameters. Because of this, the phase space of CV
is most natural for describing a phase transition. For magnetic systems,
the CV $\rhok$ are the variables associated with modes of spin-moment
density oscillations, while the order parameter is related to the
variable $\rho_0$, in which the subscript ``0'' corresponds to the peak
of the Fourier transform of the interaction potential.

The free energy of a $3D$ Ising-like system in an external field at
temperatures above $T_c$ is calculated using the non-Gaussian
spin-density fluctuations, namely the sextic measure density. The latter
is represented as an exponential function of the CV whose argument
includes the powers with the corresponding coupling constants up to the
sixth power of the variable (the $\rho^6$ model). The present publication
supplements the earlier works \cite{kpd297,p599,ykp202,ypk102}, in which
the $\rho^6$ model was used for calculating the free energy and other
thermodynamic functions of the system in the absence of an external
field. The $\rho^6$ model provides a better quantitative description of
the critical behaviour of a $3D$ Ising-like magnet than the $\rho^4$
model \cite{ykp202}.

\section{Basic relations}

We consider a $3D$ Ising-like system on a simple cubic lattice
with $N$ sites and period $c$ in a homogeneous external field $h$.
The Hamiltonian of such a system has the form
\be
H=-\half~\sli_{\ve{j},\ve{l}}\Phi(r_{\ve{j}\ve{l}})\sigma_{\ve{j}}
\sigma_{\ve{l}}-h\sli_{\ve{j}}\sigma_{\ve{j}}.
\label{ft1}
\ee
Here $r_{\ve{j}\ve{l}}$ is the distance between particles at sites
$\ve{j}$ and $\ve{l}$, and $\sigma_{\ve{j}}$ is the operator of the $z$
component of spin at the $\ve{j}$th site, having two eigenvalues +1
and -1. The interaction potential is an exponentially decreasing function
$\Phi(r_{\ve{j}\ve{l}})=A\exp\lp-r_{\ve{j}\ve{l}}/b\rp$, where
$A$ is a constant and $b$ is the radius of effective interaction. For
the Fourier transform of the interaction potential, we use the following
approximation \cite{ymo287,ykp202,ypk102}:
\be
\tPhik=\lb
\begin{array}{ll}
\tPhia(1-2b^2k^2), & k\leq B', \\
0, & B'<k\leq B.
\end{array}
\rd
\label{ft3}
\ee
Here $B$ is the boundary of the Brillouin half-zone ($B=\pi/c$),
$B'=(b\sqrt{2})^{-1}$, $\tPhia=8\pi A(b/c)^3$.

The integration over the zeroth, first, second, $\ldots$, $n$th layers
of the CV phase space \cite{ymo287,rev9789,ykpmo101,ykp202} leads to
the representation of the partition function of the system in the form
of a product of the partial partition functions $Q_n$ of individual
layers and the integral of the ``smoothed'' effective measure density
\be
Z=2^N2^{(N_{n+1}-1)/2}Q_0Q_1\cdots Q_n[Q(P_n)]^{N_{n+1}}
\int\cW_6^{(n+1)}(\rho)~({\rm d}\rho)^{N_{n+1}}.
\label{ft7}
\ee
The expressions for $Q_n$, $Q(P_n)$ are presented in
\cite{kpd297,p599,ykp202,ypk102}, and $N_{n+1}=N's^{-d(n+1)}$ ($s$ is
the RG parameter), $N'=Ns_0^{-d}$ ($d=3$ is the space dimension),
$s_0=B/B'=\pi\sqrt{2} b/c$. The sextic measure density of the ($n+1$)th
block structure $\cW_6^{(n+1)}(\rho)$ has the form
\bea
\cW_6^{(n+1)}(\rho)&=&\exp\lbr-a_1^{(n+1)}N_{n+1}^{1/2}\rho_0-
\half~\sli_{{\ve{k}}\atop{k\leq B_{n+1}}}d_{n+1}(k)\rhok\rhomk\rd\non
& & \ld-~\sli_{l=2}^3\frac{a_{2l}^{(n+1)}}{(2l)!N_{n+1}^{l-1}}~
\sli_{{\ve{k}_1,\ldots,\ve{k}_{2l}}\atop{k_i\leq B_{n+1}}}
\rho_{\ve{k}_1}\cdots\rho_{\ve{k}_{2l}}~\delta_{\ve{k}_1+\cdots+
\ve{k}_{2l}}\rbr,
\label{ft8}
\eea
where $B_{n+1}=B's^{-(n+1)}$, $d_{n+1}(k)=a_2^{(n+1)}-\btPhik$,
$\beta=1/(kT)$ is the inverse thermodynamic temperature,
$a_1^{(n+1)}$ and $a_{2l}^{(n+1)}$ are the renormalized values of the
initial coefficients after integration over $n+1$ layers of
the phase space of CV. The coefficients $a_1^{(n)}=s^{-n}t_n$,
$d_n(0)=s^{-2n}r_n$ [appearing in $d_n(k)=d_n(0)+2\btPhia b^2k^2$],
$a_4^{(n)}=s^{-4n}u_n$ and $a_6^{(n)}=s^{-6n}w_n$ are connected with the
coefficients of the ($n+1$)th layer through the recurrence relations (RR)
\bea
& & t_{n+1}=s^{(d+2)/2} t_n,\non
& & r_{n+1}=s^2\lbr -q+u_n^{1/2}Y(h_n,\alpha_n)\rbr,\non
& & u_{n+1}=s^{4-d} u_n B(h_n,\alpha_n),\non
& & w_{n+1}=s^{6-2d} u_n^{3/2}D(h_n,\alpha_n)
\label{ft9}
\eea
whose solutions
\bea
& & t_n=t^{(0)}-s_0^{d/2}h'E_1^n,\non
& & r_n=r^{(0)}+c_1E_2^n+c_2w_{12}^{(0)}(u^{(0)})^{-1/2}E_3^n+
c_3w_{13}^{(0)}(u^{(0)})^{-1}E_4^n,\non
& & u_n=u^{(0)}+c_1w_{21}^{(0)}(u^{(0)})^{1/2}E_2^n+c_2E_3^n+
c_3w_{23}^{(0)}(u^{(0)})^{-1/2}E_4^n,\non
& & w_n=w^{(0)}+c_1w_{31}^{(0)}u^{(0)}E_2^n+
c_2w_{32}^{(0)}(u^{(0)})^{1/2}E_3^n+c_3E_4^n
\label{ft10}
\eea
in the region of the critical regime are used for calculating the free
energy of the system. Here
\bea
& & Y(h_n,\alpha_n)=s^{d/2}F_2(\eta_n,\xi_n)
\lbr C(h_n,\alpha_n)\rbr^{-1/2},\non
& & B(h_n,\alpha_n)=s^{2d}C(\eta_n,\xi_n)
\lbr C(h_n,\alpha_n)\rbr^{-1},\non
& & D(h_n,\alpha_n)=s^{7d/2}N(\eta_n,\xi_n)
\lbr C(h_n,\alpha_n)\rbr^{-3/2}.
\label{ft11}
\eea
The quantity $q=\bar q\btPhia$ determines the average value of the
Fourier transform of the potential
$\beta\tilde{\Phi}(B_{n+1},B_n)=\btPhia-q/s^{2n}$ in
the $n$th layer (in this article, $\bar q=(1+s^{-2})/2$ corresponds to
the arithmetic mean value of $k^2$ on the interval $(1/s,1]$). The basic
arguments $h_n$ and $\alpha_n$ are determined by the coefficients of the
sextic measure density of the $n$th block structure. The intermediate
variables $\eta_n$ and $\xi_n$ are functions of $h_n$ and $\alpha_n$. The
expressions for both basic and intermediate arguments as well as the
special functions appearing in equations (\ref{ft11}) are the same as in
the absence of an external field (see \cite{kpd297,p599,ykp202,ypk102}).
The quantities $E_l$ in equations (\ref{ft10}) are the eigenvalues of
the matrix of the RG linear transformation. We have $E_1=s^{(d+2)/2}$.
Other the eigenvalues $E_2$, $E_3$ and $E_4$ coincide, respectively, with
the eigenvalues $E_1$, $E_2$ and $E_3$ obtained in the case of $h=0$.
The quantities $f_0$, $\varphi_0$ and $\psi_0$ characterizing the
fixed-point coordinates ($t^{(0)}=0$, $r^{(0)}=-f_0\btPhia$,
$u^{(0)}=\varphi_0(\btPhia)^2$, $w^{(0)}=\psi_0(\btPhia)^3$)
as well as the remaining coefficients in equations (\ref{ft10}) are also
defined on the basis of expressions corresponding to a zero
external field.

\section{Free energy of the system at $T>T_c$ as function of
temperature, field and microscopic parameters}

The basic idea of the free-energy calculation on the microscopic level
consists in the separate inclusion of the contributions from
short-wave (the region of the critical regime) and long-wave (the region
of the limiting Gaussian regime) modes of spin-moment density
oscillations \cite{ymo287,rev9789,ykpmo101}. The contributions
from short- and long-wave modes to the free energy $F=-kT\ln Z$
in the presence of an external field are calculated in the
$\rho^6$ model approximation according to the scheme proposed in
\cite{kpd297,p599,ykp202,ypk102}. Short-wave modes are characterized by
a RG symmetry and are described by the non-Gaussian measure density. The
calculation of the contribution from long-wave modes is based on using
the Gaussian measure density as the basis one. Here, we have developed
a direct method of calculations with the results obtained by taking into
account the short-wave modes as initial parameters.

The main peculiar feature of the present calculations lies in
using the generalized point of exit of the system from the critical
regime of order-parameter fluctuations. The inclusion of the more
complicated expression for the exit point (as a function of both the
temperature and field variables)
\be
\np=-\frac{\ln(\hhcqp)}{2\ln E_1}-1
\label{ft17}
\ee
leads to the distinction between formula for the free energy of the
system and the analogous relation at $h=0$ \cite{p599,ykp202}.
The quantity $\tilde h=h'/f_0$ is determined by the dimensionless field
$h'=\beta h$, while the quantity $\tilde h_c=\tilde\tau^{p_0}$ is
a function of the reduced temperature $\tau=(T-T_c)/T_c$. Here
$\tilde\tau=\tilde c_1^{(0)}\tau/f_0$, $p_0=\ln E_1/\ln E_2=(d+2)\nu/2$,
$\tilde c_1^{(0)}$ characterizes the coefficient $c_1$ in
solutions (\ref{ft10}) of RR, $\nu=\ln s/ln E_2$ is the critical exponent
of the correlation length. At $h=0$, $\np$ becomes
$\mt=-\ln\tilde\tau/\ln E_2-1$ (see \cite{ykpmo101,p599,ykp202}).
At $T=T_c$ ($\tau=0$), the quantity $\np$ coincides with the exit point
$n_h=-\ln\tilde h/\ln E_1-1$ \cite{kpp506}. The limiting value of the
field $\tilde h_c$ is obtained by the equality of the exit points defined
by the temperature and by the field ($\mt=n_h$).

\begin{figure}[htbp]
\centering \includegraphics[width=0.60\textwidth]{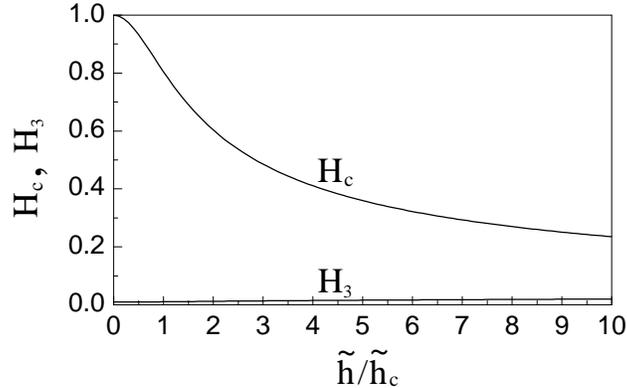}
\caption{Behaviour of quantities $H_c$ and $H_3$ with
increasing ratio $\tilde h/\tilde h_c$ for the RG parameter
$s=s^*=2.7349$ and the reduced temperature $\tau=10^{-4}$.}
\label{fig1}
\end{figure}

Having expression (\ref{ft17}) for $\np$, we arrive at the relations
\bea
& & E_1^{\np+1}=(\hhcqp)^{-1/2}, \quad
\tilde\tau E_2^{\np+1}=H_c, \quad
H_c=\tilde h_c^{1/p_0}(\hhcqp)^{-1/(2p_0)},\non
& & E_3^{\np+1}=H_3, \qquad
H_3=(\hhcqp)^{\Delb/(2p_0)},\non
& & E_4^{\np+1}=H_4, \qquad
H_4=(\hhcqp)^{\Delta_2/(2p_0)},\non
& & s^{-(\np+1)}=(\hhcqp)^{1/(d+2)},
\label{ft18}
\eea
where $\Delb=-\ln E_3/\ln E_2$ and $\Delta_2=-\ln E_4/\ln E_2$.
In contrast to $H_c$, the quantity $H_3$ takes on small values with the
variation of the field $\tilde h$ (see figure~\ref{fig1}).
The quantity $H_c$ at $\tilde h\rightarrow 0$ and near $\tilde h_c$ is
close to unity and series expansions in $H_c$ are not effective here.
The value of $s=s^*=2.7349$ (see figure caption) corresponds to
nullifying the average value of the coefficient in the term with
the second power in the effective density of measure at
the fixed point \cite{ykpmo101,p599,ykp202}.

We perform the calculations with the help of equations
(\ref{ft18}), which are valid in the general case for the regions of
small, intermediate and large field values. The inclusion of
$E_3^{\np+1}$ (or $H_3$) leads to the formation of the first confluent
corrections in the expressions for thermodynamic characteristics of the
system. The quantity $E_4^{\np+1}$ (or $H_4$) is responsible for the
emergence of the second confluent corrections. We disregard the second
confluent correction in our calculations. This is due to the fact that
the contribution from the first confluent correction to thermodynamic
functions near the critical point ($\tau=0$, $h=0$) is more significant
than the small contribution from the second correction ($\hhcqp\ll 1$,
$\Delb$ is of the order of 0.5 and $\Delta_2 > 2$, see \cite{kpd297}).

Collecting the contributions from all regimes of fluctuations at
$T>T_c$ in the presence of an external field and using the relation
for $s^{-(\np+1)}$ from equations (\ref{ft18}), we obtain the following
expression for the total free energy of the system:
\bea
F&=&-kTN\Biggl[ \gamma'_0+\gamma'_1\tau+\gamma'_2\tau^2+
(\bar \gamma_3^{(0)+}+\bar \gamma_3^{(1)+}\ccaaHd)(\hhcqp)^{3/5}\non
& & +\frac{\bar \gamma_4^+(h')^2}{\btPhia}~(1-\bar g_1\ccaaHd)
(\hhcqp)^{-2/5}\Biggr].
\label{ft52}
\eea
The advantage of the method presented in this article is the possibility
of deriving analytic expressions for the free-energy coefficients as
functions of the microscopic parameters of the system (the lattice
constant $c$ and parameters of the interaction potential, i.e. the
effective radius $b$ of the potential, the Fourier transform $\tPhia$ of
the potential for $k=0$). The coefficients of the non-analytic component
of the free energy depend on $H_c$. The terms proportional to $H_3$ in
equation (\ref{ft52}) determine the confluent corrections by the
temperature and field. As is seen from the expression for $F$, the free
energy of the system at $\tilde h=0$ and $\tilde\tau=0$, in addition to
terms proportional to $\tilde\tau^{3\nu}$ (or $\tilde h_c^{6/5}$) and
$\tilde h^{6/5}$, contains the terms proportional to
$\tilde\tau^{3\nu+\Delb}$ and $\tilde h^{6/5+\Delb/p_0}$, respectively.
At $\tilde h\neq 0$ and $\tilde\tau\neq 0$, the terms of both types are
present.

\section{Conclusions}

An analytic method for calculating the total free energy of a $3D$
Ising-like system near the critical point is developed on the microscopic
level in the higher non-Gaussian approximation based on the sextic
distribution for modes of spin-moment density oscillations (the $\rho^6$
model). The simultaneous effect of the temperature and field on the
behaviour of the system is taken into account. An external field is
introduced in the Hamiltonian of the system from the outset. In contrast
to previous studies on the basis of the asymmetric $\rho^4$ model
\cite{kpp506,pkp105,kpp406,p806}, the field in the initial process of
calculating the partition function of the system is not included in the
Jacobian of transition from the set of spin variables to the set of CV.
Such an approach leads to the appearance of the first, second, fourth
and sixth powers of CV in the expression for the partition function and
allows us to simplify the mathematical description because the odd part
is represented only by the linear term.

The main distinctive feature of the proposed method is the separate
inclusion of the contributions to the free energy from the short- and
long-wave spin-density oscillation modes. The generalized point of exit
of the system from the critical regime contains both the temperature and
field variables. The form of the temperature and field dependences for
the free energy of the system is determined by solutions of RR near the
fixed point. The expression for the free energy obtained at temperatures
$T>T_c$ without using power series in the scaling variable and without
any adjustable parameters can be employed in the field region near
$\tilde h_c$. The limiting field $\tilde h_c$ satisfies the condition of
the equality of sizes of the critical-regime region by the temperature
and field (the temperature and field effects on the system in the
vicinity of the critical point are equivalent)
\cite{kpp506,pkp105,kpp406,p806}. Proceeding from the expression
for the free energy, which involves the leading terms and terms
determining the temperature and field confluent corrections, we can find
other thermodynamic characteristics (average spin moment, susceptibility,
entropy and specific heat) by direct differentiation of equation
(\ref{ft52}) with respect to field or temperature.

\end{document}